# Defect states and their electric field-enhanced electron thermal emission in heavily Zr-doped β-Ga$_2$O$_3$ crystals

Rujun Sun[1], Yu Kee Ooi[1], Arkka Bhattacharyya[1], Muad Saleh[2], Sriram Krishnamoorthy[1], Kelvin G. Lynn[2], and Michael A. Scarpulla [1,3,*]

[1.] Electrical and Computer Engineering, University of Utah, Salt Lake City, UT, 84112, USA

[2.] Materials Science & Engineering Program, Washington State University, Pullman, WA, 99164 USA

[3.] Materials Science and Engineering, University of Utah, Salt Lake City, UT, 84112, USA

**Abstract**

Performing deep level transient spectroscopy (DLTS) on Schottky diodes, we investigated defect levels below the conduction band minima ($E_c$) in Czochralski (CZ) grown unintentionally-doped (UID) and vertical gradient freeze (VGF)-grown Zr-doped β-Ga$_2$O$_3$ crystals. In UID crystals with an electron concentration of $10^{17}$ cm$^{-3}$, we observe levels at 0.18 eV and 0.46 eV in addition to the previously reported 0.86 (E2) and 1.03 eV (E3) levels. For $10^{18}$ cm$^{-3}$ Zr-doped Ga$_2$O$_3$, signatures at 0.30 eV (E15) and 0.71 eV (E16) are present. For the highest Zr doping of 5×$10^{18}$ cm$^{-3}$, we observe only one signature at 0.59 eV. Electric field-enhanced emission rates are demonstrated via increasing the reverse bias during measurement. The 0.86 eV signature in the UID sample displays phonon-assisted tunneling enhanced thermal emission and is consistent with the widely reported E2 (Fe$_{Ga}$) defect. The 0.71 eV (E16) signature in the lower-Zr-doped crystal also exhibits phonon-assisted tunneling emission enhancement. Taking into account that the high doping in the Zr-doped diodes also increases the electric field, we propose that the 0.59 eV signature in the highest Zr-doped sample likely corresponds to the 0.71 eV signature in lower-doped samples. Our analysis highlights the importance of testing for and reporting on field-enhanced emission especially the electric field present during DLTS and other characterization experiments on β-Ga$_2$O$_3$ along with the standard emission energy, cross section, and lambda-corrected trap density. This is important because of the intended use of β-Ga$_2$O$_3$ in high-field devices and the many orders of magnitude of possible doping.

*Corresponding Author, Email: mike.scarpulla@utah.edu*

Ultra-wide bandgap β-Ga$_2$O$_3$ is promising in applications such as power electronic devices, solar-blind UV photodetectors, solar cells, and sensors due to its n-type dopability from $10^{15}$ to $10^{20}$ cm$^{-3}$ and bandgap of 4.5 to 4.8 eV depending on the optical axis and measurement methods[1,2]. Melt growth techniques, such as edge-defined film-fed growth[3] (EFG), Czochralski method[4] (CZ), floating-zone[5] (FZ), have been successfully used for large-size single crystals. Conductive β-Ga$_2$O$_3$ substrates, essential components for high breakdown voltage vertical devices, are achieved by Si[6] and Sn[4,7] doping with carrier concentration exceeding $10^{18}$ cm$^{-3}$. Elements like Zr[8], Hf[9], Nb[10], and Ta[11] have also been demonstrated as shallow, high concentration dopants in Ga$_2$O$_3$ crystals. The Zr-doped Ga$_2$O$_3$ crystals grown by Verneuil technique exhibit high conductivity with blue color[12]. Recently, larger size Zr-doped β-Ga$_2$O$_3$ crystals have been grown by vertical gradient freeze technique[8] with free carrier concentration above $10^{18}$ cm$^{-3}$. Higher shallow dopant concentration results in higher free carrier concentration but lowers the mobility due to impurity scattering[13]. Free carrier absorption is observed especially for heavily-doped Ga$_2$O$_3$ and the plasma frequency shifts to shorter wavelengths with higher free carrier concentration[14].

Deep defect states also play a critical role in electrical and optical properties, both in terms of the familiar lowering of minority carrier lifetimes via nonradiative recombination but also in terms of creating semi-insulating material. For example, in β-Ga$_2$O$_3$:Fe semi-insulating crystals, the Fermi energy is pinned at the $Fe^{3+}/Fe^{2+}$ charge transition level approximately $E_c$-0.78 eV (the E2 level from DLTS)[15] resulting in vanishingly-small carrier densities. Acceptors of Mg can also compensate intentional or unintentional shallow donors, reducing net doping and mobility [4,16]. In terms of native defects, the relaxed Ga vacancy, sometimes described as a $(2V_{Ga}+Ga_i)$ complex or split vacancy, is a compensating acceptor and has been observed directly in electron microscopy in heavily Sn-doped EFG β-$Ga_2O_3$ crystals[17]. MOVPE-grown homoepitaxial β-$Ga_2O_3$ thin films are still resistive even when they are doped to $10^{22}$ $cm^{-3}$ of Si in O-rich conditions, indicating the abundance of compensating acceptor, in which the activation energy of 0.50-0.65 eV has been characterized by thermally stimulated luminescence (TSL)[18]. While nominally, effective mass donors are interchangeable in terms of adding net shallow doping, each particular extrinsic impurity can form a unique variety of complexes with all defects present and thus each different dopant in principle can result in a different ensemble of defect states in the bandgap. Thus, it is important to characterize the defect ensembles present for different shallow dopants across available ranges of doping, as this investigation begins to do for Zr.

Deep level transient spectroscopy (DLTS) uses thermal emission from traps to probe defects within approximately 1.0 eV from the band edges and is sensitive to defect concentrations as small as ~$10^{-4}$ times the net shallow doping in typical circumstances. For defects with Coulomb or similar potentials for thermal emission, the presence of a strong electric field (F) within the depletion region lowers the apparent emission barrier via the Poole-Frenkel effect[19,20]. Phonon-assisted tunneling[21,22] emission can also increase the emission rate in high fields. Both effects typically manifest by enhancing the thermal emission rate which can be described as a lowering of the apparent thermal emission activation energy with increasing field. The emission rate for an electron occupying a trap at energy $E_t$ is given by:

$$e_n = v_{th}\sigma_n \left(N_c - n\right)\exp\left(\frac{-E}{k_B T}\right)\exp\left(\frac{\Delta E}{k_B T}\right)$$

in which $v_{th}$ is the thermal velocity, $\sigma_n$ is the capture cross section, $N_c$-n is the density of unoccupied states in the conduction band (=$N_c$ in a depletion width), $E = E_c-E_t$ and $\Delta E = const \cdot F^\gamma$. For the Poole-Frenkel effect $\gamma=2$ while for phonon-assisted tunneling $\gamma=½$. The apparent energy $E_{app}$ measured in a DLTS experiment would be $E_{app} = E_C - E_t - const \cdot F^\gamma$ and by linearly extrapolating a plot of $e_n$ vs $F^\gamma$ to F=0 one could determine the field-free value of $E_c-E_t$.

DLTS has been primarily measured on UID bulk crystals and rather lightly-doped epitaxial β-Ga$_2$O$_3$ layers for which the effects of electric field on emission rate are similar and relatively small, thus results can be compared with limited discrepancies. For example, several deep electron trap signatures have been reported in β-Ga$_2$O$_3$, e.g., Ec-0.21 eV[23] (E10), Ec-0.33 eV[23] (E11, an native defect), $E_C$-0.40 eV[24] (E9), $E_C$-0.6 eV (E1), $E_C$-0.75 eV (E2*, an native defect)[15], $E_C$-0.78 eV (E2, Fe$_{Ga}$)[15], $E_C$-0.95 eV (E3, Ti$_{Ga}$)[25], $E_C$-1.04 eV (E3) and $E_C$-1.2 eV (E4)[15,24-29]. Amongst these reported defects, the E3 state is donor-like and its apparent emission energy shifts with electric field according to the Poole-Frenkel effect[30,31]. The E2 state exhibits phonon-assisted tunneling emission enhancement[31].

In this work, we investigate deep states in heavily Zr-doped Ga$_2$O$_3$ using DLTS and demonstrate the electric field dependent emission rates. In these heavily-doped samples, the diffusion potential as well as applied bias are sufficient to induce field induced emission enhancement. In addition to displaying smaller apparent activation energies compared to those from UID samples, the Zr-doped Ga$_2$O$_3$ crystals show broadening of some DLTS trap signatures. We interpret this broadening to the emission of one trap type in the steep spatial field gradient which causes a distribution of emission rates for defects at different depths, as opposed to a real distribution of trap energies which might be the case e.g. for interface defects. By variation of the measuring bias in DLTS and modeling of the wide emission feature using the emission rate equation, we find that the 0.86 eV (E2) in the UID sample and the 0.71 eV (E16) and 0.59 eV (E16) signatures in Zr-doped samples exhibit phonon-assisted tunneling effects.

Unintentionally-doped (UID) and Zr-doped $\beta$-Ga$_2$O$_3$ were grown by the Czochralski (CZ) and vertical gradient freeze (VGF) techniques, respectively[8,9]. Heavily-doped crystals tend to grow as spirals in CZ, thus VGF was used for the Zr-doped crystals. Three types of samples were investigated, namely, UID, #1 Zr ($\sim 1 \times 10^{18}$ cm$^{-3}$), and #2 Zr ($\sim 5 \times 10^{18}$ cm$^{-3}$) where UID crystal was utilized as a control sample. Before the contact deposition, the samples were cleaned by acetone, isopropyl alcohol, and deionized water. Ni/Au Schottky contacts and Ti/Au ohmic contacts were deposited on cleaved (100) planes. As seen in Fig.1d, lateral and vertical geometries were used for Zr-doped and UID samples, respectively. *J-V* and *C-V* measurements were performed using Keithley 4200 SCS parameter analyzer. DLTS was performed using a Sula DLTS spectrometer from 80 K to 430 K (temperature recorded on the sample surface) in the dark with a 100 mV, 1 MHz AC signal. Except for the experiments used for investigating field-dependent emission rates, the fill voltage was 0 V and measure voltage -2 V for Zr-doped samples and 0 and -1 V respectively for the UID one.

We first tested the general characteristic of the Schottky junctions before DLTS measurements. **Fig. 1a** shows the *J-V* curves of UID, #1, and #2 Zr-doped Ga$_2$O$_3$ Schottky diodes. UID and #1 samples show leakage current less than $10^{-7}$ A/cm$^2$ from 0 V to -5.0 V. A notable leakage current is observed under reverse bias in the higher-doped Zr sample #2, which is likely caused by tunneling and image force lowering over Schottky barrier. The ideality factor n is defined as n =

$\frac{q}{k_\text{B}T\frac{\text{d}}{\text{d}V}\ln(J)}$ using the thermionic emission (TE) model and was determined over a forward voltage ($V_\text{f}$) range from 0.50 to 0.80 V ($V_\text{f} \gg k_\text{B}T/q$), where $q$ is the electron charge, $k_\text{B}$ is Boltzmann constant, $T$ is temperature, $J$ is current density, $V$ is applied voltage. The ideality factors for the representative UID, #1, and #2 Zr-doped Ga$_2$O$_3$ Schottky diodes are 1.11, 1.09, and 1.37 respectively. The increased ideality factor in the #2 Zr sample indicates some mixed thermionic and field emission instead of a pure TE model. The $\omega C_p/G_p$ values derived from C-V measurements for all samples are much greater than 5 for all biases including near -5 V near 390 K (**Fig.1b**), indicating the reliability of the capacitance measurements including DLTS[32]. Their leakage current values are also within the tolerable current range (150 µA) of the Sula system for performing DLTS measurement. Profiles of doping concentration as a function of depletion width calculated from C-V data are shown in **Fig. 1c**. The net doping concentrations for UID, #1 and #2 Zr-doped Ga$_2$O$_3$ are $1.8\times10^{17}$ cm$^{-3}$, $1.5\times10^{18}$ cm$^{-3}$, and $5.0\times10^{18}$ cm$^{-3}$, respectively. In **Fig.1d**, the built-in voltages ($V_\text{bi}$) derived from Mott-Schottky analysis of $1/C^2$ vs voltage curves for the UID, #1 Zr and #2 Zr samples are 1.38, 1.58, and 1.23 V, respectively. Ideally, the $V_\text{bi}$ should increase at 60 mV per decade of net doping; our results indicate that some non-ideal factors help to determine the V$_\text{bi}$.

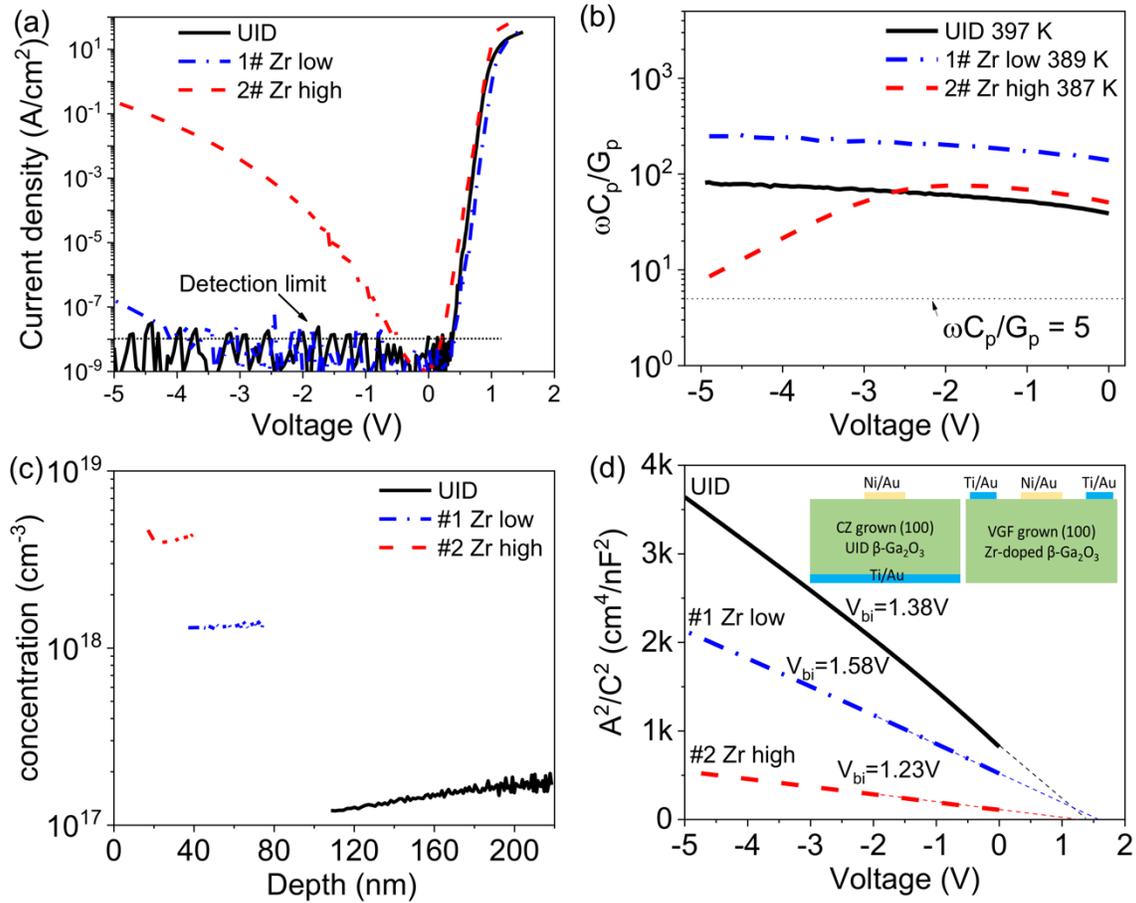

Figure 1. (a) J-V at 310 K, (b) $\omega C_p/G_p$ from *C-V* characteristics for UID and Zr-doped β-Ga$_2$O$_3$ Schottky diodes at 390 K, (c) carrier density vs depletion width from zero to -5.0 V at 310 K, (d) $V_{bi}$ derived from A$^2$/C$^2$ vs voltage from -2 V to 0 V at 310 K and the schematics of Schottky diodes are included.

The activation energy, electron capture cross-section, and number density of deep defects can be extracted from conventional DLTS measurements on these n-type Schottky diodes. The voltage is pulsed to a less-negative or 0 bias during the filling pulse and held at a reverse bias during the measurement of emission. At steady state near equilibrium, the thermal emission rate from an electron trap is $e_n = \frac{g_0}{g_1}\sigma_n v_n^{th} N_C \exp\left(-\frac{E_{app}}{k_B T}\right)$ derived from the principle of detailed balance[33], in which $\sigma_n$ is the electron capture cross-section, $v_n^{th}$ is the thermal velocity, $N_C$ is the conduction band effective density of states, and $\frac{g_0}{g_1}$ is the ratio of electronic degeneracies of empty and occupied deep states, which is taken as 1.0 by convention to obtain an apparent cross-section. For a concentration of traps small compared to the net doping and uniform doping and defects vs depth, the trap concentration is determined using $N_T = 2N_d \Delta C/C_{rb} \{W_r^2/((W_r - \lambda)^2 - (W_0 - \lambda)^2))\}$ in which $C_{rb}$ is the limiting capacitance under reverse bias, $\Delta C$ is the magnitude of the emission capacitance transient, and the term in curly brackets is the so-called lambda correction in which $W_r$ and $W_0$ are the depletion widths at the pulse and reverse biases, respectively, and $\lambda = \sqrt{\frac{2\varepsilon_0 \varepsilon_r}{q^2 N_d}}(E_F - E_T)$ for a trap with energy $E_T$[34].

Representative DLTS spectra of UID, #1 and #2 Zr-doped β-Ga$_2$O$_3$ are shown in **Fig.2a**. The labeling of defect signatures up to 10 (13 including another 3[23]) follow the designations from Ref. [35] and we add E14-16 here in this work. The UID sample shows four majority defects located at 0.18 eV (E14), 0.46 eV (E12), 0.86 eV (E2), and 1.03 eV (E3) below $E_C$, respectively. Their corresponding cross-sections $\sigma_n$ are 7.7×10$^{-16}$, 1.1×10$^{-14}$, 1.1×10$^{-13}$, and 8.6×10$^{-13}$ cm$^2$. For the sample Zr-doped to 10$^{18}$ cm$^{-3}$, a signature E15 at 0.30 eV (σ=2.8×10$^{-17}$ cm$^2$) is observed coexisting with a dominant 0.71 eV (E16) signature (σ=2.6×10$^{-15}$ cm$^2$). When the concentration of Zr is increased to 5×10$^{18}$ cm$^{-3}$, only one defect at 0.59 eV (E16) (σ=7.5×10$^{-15}$ cm$^2$) is observed. The signature of $E$c-0.86 eV from UID has the highest concentration of 4.1×10$^{16}$ cm$^{-3}$ (Table 1), which is at the same order magnitude of the reported highest defect density from E2 and E3 defects in other UID crystals by CZ and EFG methods[26,27]. Note that the broad DLTS signal in Zr-doped samples is broader than what would be predicted from a single $E_T$ value using the emission rate equation and rate window analysis[36]. An additional feature in the data is a broad positive ΔC signals (as opposed to the negative signals for the traps discussed so far) are observed from the Zr-doped samples above 340 K and are more pronounced for the sample with higher Zr concentration. A change in polarity of the DLTS transient conventional DLTS could arise from a majority capture transient or possibly from minority emission[36] in the bulk layer as well as from artifacts such as high series resistance (high $R_s$ is not our case)[37]. We are dubious of the possibility of minority carrier emission transients since these are Schottky not p-n junctions and no above gap light was present. The origins of these signals are the subject of further investigations, however the broad and asymmetric DLTS signals vs. temperature are consistent with a surface defect origin[34]. This is further supported by the DLTS signals pulsing from -6 V to -4 V which probes a deeper depth from the surface (>50 nm) in #1 Zr sample in Fig.2c for which the positive transient becomes smaller. Since the depletion depths for #1 and #2 Zr-doped samples at -2 V are 30 nm and 15 nm respectively, the occupation could change under filling bias if localized charges exist near the surface due to possibly un-optimized surface preparation.

For negative majority emission transients, the energy levels obtained from DLTS measurements can vary depending on the electrical field in which the defect sits during the measurement phase of the experiment which in turn changes as a function of built-in and applied bias and doping density[20,30,34,38]. As shown in Fig.2b, the DLTS signals in the UID sample shift toward lower temperature and the asymmetric tails at lower temperature side enlarge for increasing reverse biases – effects which are more pronounced for the #1 Zr doped sample (Fig.2c). Note the E14 signature in UID not only shifts but also seems to split into two peaks (Fig.S1). These suggest E14 might consist of two defect signatures, and their DLTS shapes shift at different rates under field. Poole-Frenkel and tunneling models are two primary mechanisms for the enhancement of emission rates, which can be distinguished by plotting the logarithm of emission rate against the square root of the electric field or the square of the electric field, respectively[21]. The Poole-Frenkel effect results in moderate enhancement of emission rate and typically applies for the deep donor, while the phonon-assisted tunneling is possible for all charge states in a strong field and does not in general require a Coulomb potential. The electric field dependent emission rate can be investigated by double-correlation DLTS and isothermal capacitance transient analysis[19] which are unfortunately outside of our instrument's ability. We attempted to model the broad DLTS signatures using the superposition of multiple defect signatures with different weightings, by applying the same rate windows as used in measurements[39]. The physical motivation here is trying to capture the fact that defects in different locations in the depletion width would experience different fields which in turn would result in different emission energies for the same defect type as a function of depth. Tentatively, we simulate the corresponding DLTS spectra with the minimum possible number of $E_T$ values (we found 3 values captured most of the behavior without having to introduce a continuous distribution) each sharing the same σ extracted from the peak maximum from Fig.2a (see Supplementary Materials). Among the inputs $E_T$ (actually Et+ΔE), the smallest $E_T$ value corresponds to the largest emission rate which relates to F field at the left boundary of $x$ in Fig.2d. Fig.2d shows the electric field $F(x) = \sqrt{\frac{2qN_d(V_{bi}-V)}{\varepsilon_0\varepsilon_r}} - \frac{qN_d}{\varepsilon_0\varepsilon_r}x$ at the locations that electron occupation at defect changes under emptying bias. The left boundary of $x$ value is at ($W_0$-λ) depth where each trap level crosses the Fermi level under 0 V, and then we can calculate the field at that point. We recalculated the emission rate using the smallest $E_T$ with its σ at a specific temperature and check the relationship between $e_n$ and F. Fig.2e shows the emission rate for the E2 signature in UID at 375 K vs maximal F where electron occupation at defect changes. It turns out to follow the square of F, indicating emission rate enhanced by phonon-assisted tunneling mode.

Interestingly, the 0.71 eV signature in low Zr doped sample #1 and the 0.59 eV signature in high Zr doped sample #2 also follow the $E^2$ rule (Fig.2f), indicating that they may arise from the same defect emitting at different rates by virtue of the different F fields in the two samples. The fact that the extracted cross-sections are also close suggests this possibility. Thus both are labeled as E16 signature. The Fig.2f provides the best evidence that the defect seen in Zr#2 is the same defect seen in Zr#1. The simulated results for the E14 and E12 signatures in UID and E15 in #1 Zr are hard to gain good agreement with the measured signals due to the existence of the long asymmetry tail under the high electric field, suggesting a different enhanced mechanism.

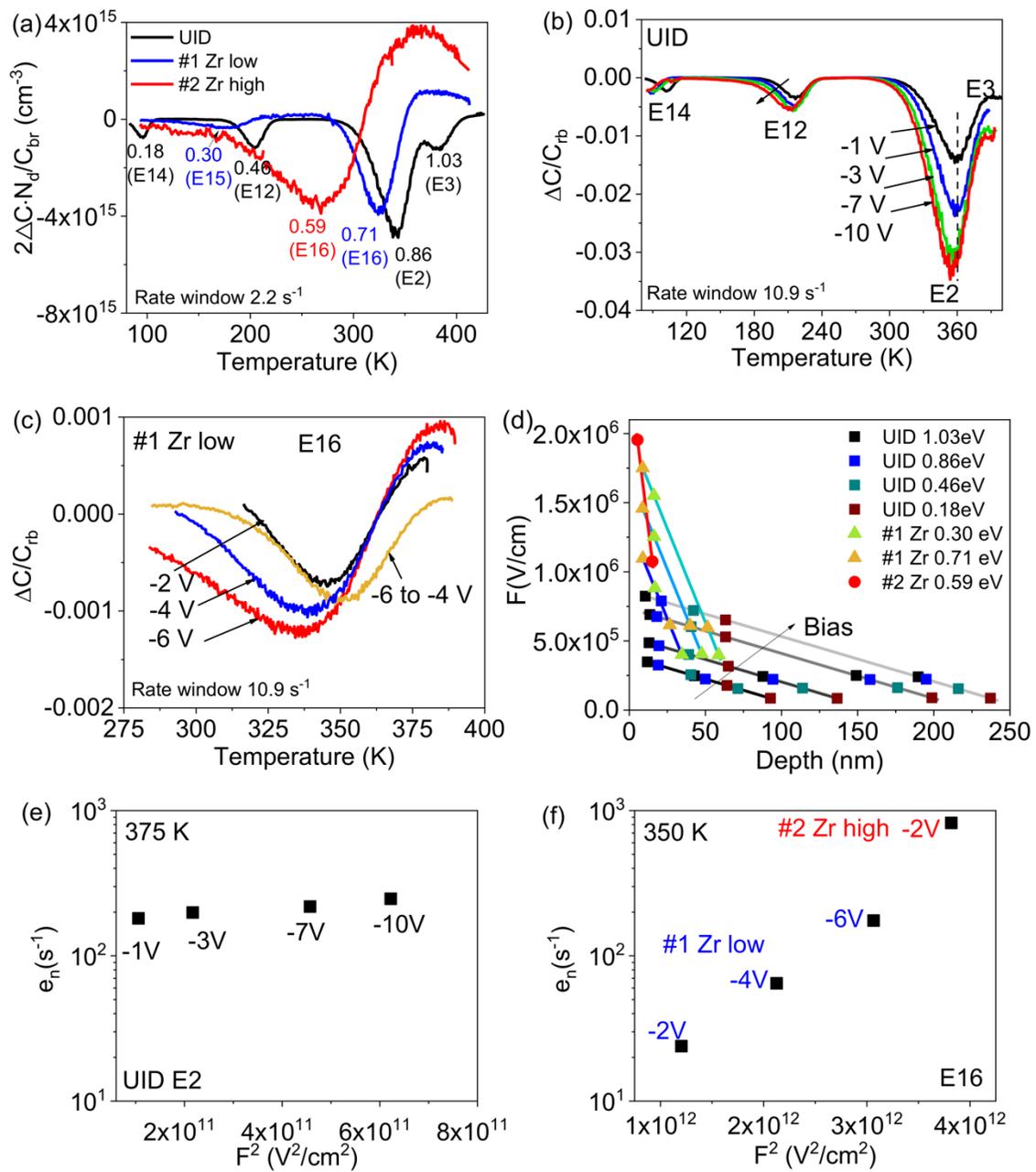

Figure 2. (a) representative DLTS spectra of devices from UID, #1, and #2 Zr-doped β-$Ga_2O_3$ crystals. DLTS spectra measured with different noted reverse biases pulsed to 0 V for (b) UID and (c) #1 Zr doped crystals (except -6 to -4 V as noted). Other peaks behaved similarly in the Zr doped samples but are not shown here. (d) electric field as a function of depth electron occupation

changes for different samples and under the applied reverse biases. The colorful points at left and right for each defect represent electric fields at positions of the ($W_0$-λ) and ($W_r$-λ), respectively, at t=0 under reverse bias. (e) the maximal emission rate vs calculated local maximum electric field for E2 in UID, and (f) 0.71 eV in #1 Zr doped sample and 0.59 eV in #2 Zr doped sample. For Zr-doped samples, positive transient is subtracted for simulating.

We also plot our emission rate data along with those from prior reports in terms of $\ln(\frac{e_n}{T^2})$ vs. $1/T$ in **Fig. 3**. For the UID crystal, the signature at 1.03 eV with cross-section of 8.6×10$^{-13}$ cm$^2$ agrees well with the E3 in EFG and CZ UID crystals[27] and is likely from native defects since its concentration has been observed to increase after neutron irradiation[40] but we cannot rule out the possibility of Ti origin yet[25]. The signature we observe at 0.86 eV is consistent with E2 which has been attributed to the Fe$_{Ga}$ 3$^+$/2$^+$ acceptor charge transition level. The dependence of emission on F field in this paper and Polyakov's work[31] also support this conclusion. The signatures at 0.46 eV (E12) and 0.18 eV (E14) have not been reported previously in CZ UID crystals. The 0.46 eV signature is consistent with a defect level at $E_c$-0.42 eV (E12) in plasma-assisted molecular beam epitaxy (PAMBE) Ge-doped films[23] considering the similar doping, bias conditions, and measured capture cross-sections. Due to the high purity in PAMBE growth, it might be from a native defect instead of an impurity, however this remains to be proven. The E14 signature we observe at 0.18 eV appears similar to a defect signature at $E_c$-0.21 eV reported from Ge-doped PAMBE-grown samples considering $E_T$, σ and F-field values, however more work is needed to confirm this.

In the Zr doped samples, the apparent $E_T$ is lowered by the electric field, thus cannot be directly compared with the reported $E_T$ from UID or lightly-doped samples. Considering one order of magnitude uncertainty for σ and the lowering of apparent activation energy, the measured 0.71 eV from #1 and 0.59 eV from #2 of Zr-doped crystals likely correspond to the reported 0.75 eV (E2*) in UID crystals. Table 1 summarizes the measured defect levels, their behaviors under varying electric field, and comparisons with reported defects. Given the complexity of defect emission, more work is needed to reveal the charge state and donor/acceptor nature of measured defects by DLTS. However, our analysis points out the importance of reporting DLTS signatures considering electric field enhanced emission rate for β-$Ga_2O_3$ due to many orders of magnitude of possible doping and the expected applications in high-field applications.

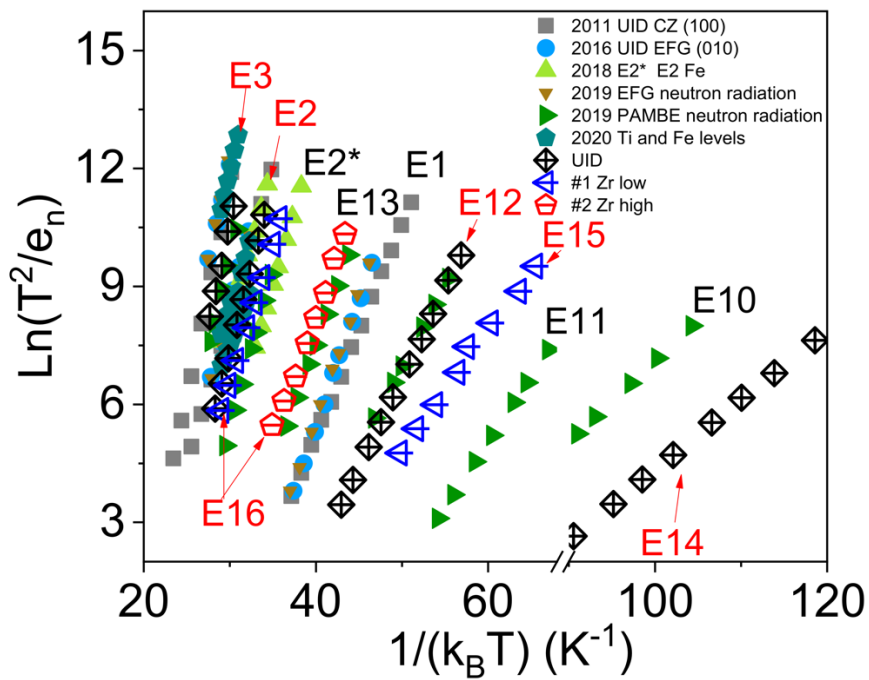

Figure 3. Arrhenius plot of each detected defect by this work and from literatures[15,23,25-27]

Table 1. Summary of defect levels below $E_C$, its behaviors under varying electric field and comparisons with reported defects. We have introduced the E14 – E16 signatures herein as new because they do not correspond to previously-reported DLTS signatures.

| Samples | $E_T$ (eV) | $\sigma$ (cm$^2$) | $N_T$ (cm$^{-3}$) | Notation | Notes |
|---|---|---|---|---|---|
| UID | 0.18 | 7.7×10$^{-16}$ | 2.2×10$^{15}$ | E14 | shift and split under higher F |
| | 0.46 | 1.1×10$^{-14}$ | 5.0×10$^{15}$ | E12 | asymmetrically-broadened towards lower temperatures under higher F |
| | 0.86 | 1.1×10$^{-13}$ | 4.1×10$^{16}$ | E2 | Phonon assisted tunneling |
| | 1.03 | 8.6×10$^{-13}$ | 1.5×10$^{16}$ | E3 | |
| #1 Zr low | 0.30 | 2.8×10$^{-17}$ | 1.2×10$^{15}$ | E15 | asymmetrically-broadened towards lower temperatures under higher F |
| | 0.71 | 2.6×10$^{-15}$ | 1.6×10$^{16}$ | E16 | Phonon assisted tunneling |
| #2 Zr high | 0.59 | 1.2×10$^{-15}$ | 1.2×10$^{16}$ | E16 | |

In conclusion, Zr-doped and UID β-Ga$_2$O$_3$ crystals grown by the VGF and CZ methods were investigated by DLTS. The apparent electron emission activation energies of the CZ UID crystal were 0.18 (E14), 0.46 (E12), 0.86 (E2), and 1.03 eV (E3) below $E_C$. For the 1# low-Zr-doped Ga$_2$O$_3$, apparent energies of 0.30 eV (E15) and 0.71 eV (E16) were present. For the #2 higher-Zr-doped sample, only a signature at 0.59 eV (E16) was observed. Electric-field-dependent emission rates were demonstrated by varying the reverse bias and showing the consistency of the measured defect energies for differently-doped Zr samples. The 0.86 eV (E2) in UID exhibits emission field enhancement consistent with phonon-assisted tunneling. The signatures observed at 0.71 eV and This manuscript is accepted to publish in Applied physics letters on Nov 5 2020.0.59 eV in the two differently-Zr-doped crystals are assigned to the same E16 signature after demonstrating that the combination of applied and built-in fields with phonon-assisted tunneling enhanced emission can account for their different apparent emission energies.

See supplementary material for DLTS spectra of UID sample in 80-110K range (E14 signature) under different reverse biases, and simulated spectra of E2 and E12 in UID as examples to demonstrate how the DLTS simulation performs in this work.

This material is based upon work supported by the Air Force Office of Scientific Research under award number FA9550-18-1-0507 (Program Manager: Dr. Ali Sayir). Any opinions, findings, conclusions, or recommendations expressed in this material are those of the author(s) and do not necessarily reflect the views of the United States Air Force. This work is dedicated to the memory of the late Prof. Kelvin G. Lynn.

**Data Availability**

The data that support the findings of this study are available from the corresponding author upon reasonable request.

This manuscript is accepted to publish in Applied physics letters on Nov 5 2020.30   M. E. Ingebrigtsen, A. Yu Kuznetsov, B. G. Svensson, G. Alfieri, A. Mihaila, and L. Vines,  J. Appl. Phys. **125** (18), 185706 (2019).
31   A. Y. Polyakov, In-Hwan Lee, N. B. Smirnov, I. V. Shchemerov, A. A. Vasilev, A. V. Chernykh, and S. J. Pearton,  J. Phys. D: Appl. Phys. **53** (30), 304001 (2020).
32   Dieter K. Schroder, *Semiconductor Material and Device Characterization, 3rd Edition*. (2005), p.840.
33   W. Shockley and W. T. Read,  Phys. Rev. **87** (5), 835 (1952).
34   Peter Blood and J. W. Orton, *The electrical characterization of semiconductors: majority carriers and electron states*. (Academic Press, London ; San Diego, Calif., 1992), pp.xxiii.
35   Wang Zhengpeng, Chen Xuanhu, Ren Fangfang, Gu Shulin, and Ye Jiandong,  J. Phys. D: Appl. Phys. (2020).
36   D. V. Lang,  J. Appl. Phys. **45** (7), 3023 (1974).
37   A. Broniatowski, A. Blosse, P. C. Srivastava, and J. C. Bourgoin,  J. Appl. Phys. **54** (6), 2907 (1983).
38   Oleg Mitrofanov and Michael Manfra,  J. Appl. Phys. **95** (11), 6414 (2004).
39   Artur Scheinemann and Andreas Schenk,  physica status solidi (a) **211** (1), 136 (2014).
40   Esmat Farzana, Max F. Chaiken, Thomas E. Blue, Aaron R. Arehart, and Steven A. Ringel,  APL Mater. **7** (2), 022502 (2019).
18